\documentclass[aps,pra,twocolumn,groupedaddress,showpacs]{revtex4}
\usepackage{graphicx}
\usepackage{amssymb}

\begin{document}


\title{Photoassociation and coherent transient dynamics in the interaction of ultracold rubidium atoms with shaped femtosecond pulses - I. Experiment}



\author{Wenzel Salzmann}
\altaffiliation{present address: Fraunhoferinstitut f{\"u}r Physikalische
  Messtechnik IPM, Heidenhofstr. 8, 79110 Freiburg, Germany}
\author{Terry Mullins}
\author{Simone G{\"o}tz}
\author{Magnus Albert}
\altaffiliation{present address: Department of Physics, University of Aarhus,
  Ny Munkegade, 8000 Aarhus, Denmark}
\author{Judith Eng}
\author{Roland Wester}
\author{Matthias Weidem{\"u}ller}
\altaffiliation{present address: Physikalisches Institut, Universit{\"a}t
  Heidelberg, Philosophenweg 12, 69120 Heidelberg, Germany}
\affiliation{Physikalisches Institut, Universit\"at Freiburg, Hermann Herder
  Str. 3, 79104 Freiburg, Germany}
\email[]{weidemueller@physi.uni-heidelberg.de}

\author{Fabian Weise}
\author{Andrea Merli}
\author{Stefan M. Weber}
\author{Franziska Sauer}
\author{Ludger W\"oste}
\author{Albrecht Lindinger}
\email[]{lindin@physik.fu-berlin.de}

\affiliation{Institut f\"ur Experimentalphysik, Freie Universit\"at
Berlin, Arnimallee 14, 14195 Berlin, Germany}

\begin{abstract}
We experimentally investigate various processes present in the
photoassociative interaction of an ultracold atomic sample with
shaped femtosecond laser pulses. We demonstrate the photoassociation
of pairs of rubidium atoms into electronically excited, bound
molecular states using spectrally cut femtosecond laser pulses tuned
below the rubidium D1 or D2 asymptote. Time-resolved pump-probe
spectra reveal coherent oscillations of the molecular formation
rate, which are due to coherent transient dynamics in the electronic
excitation. The oscillation frequency corresponds to the detuning of
the spectral cut position to the asymptotic transition frequency of
the rubidium D1 or D2 lines, respectively. Measurements of the
molecular photoassociation signal as a function of the pulse energy
reveal a non-linear dependence and indicate a non-perturbative
excitation process. Chirping the association laser pulse allowed us
to change the phase of the coherent transients. Furthermore, a
signature for molecules in the electronic ground state is found,
which is attributed to molecule formation by femtosecond
photoassociation followed by spontaneous decay. In a subsequent
article [A. Merli {\it et al.}, submitted] quantum mechanical
calculations are presented, which compare well with the experimental
data and reveal further details about the observed coherent
transient dynamics.
\end{abstract}

\maketitle

\section{Introduction}

Since the first proposal of light-induced formation of ultracold
molecules by Thorsheim {\it et al.} \cite{thorsheim1987} and its
experimental realization by Lett {\it et al.} \cite{lett1993}, a
rapidly progressing field has developed for the creation and
manipulation of ultracold molecular gases by photoassociation
\cite{MasnouReview2001,jones2006} and magnetoassociation using
Feshbach resonances \cite{jochim2003,timmermans1999}. On the way to
lowest energies in both the external and internal degrees of
freedom, numerous techniques have been developed to produce large
samples of molecular quantum gases in their absolute ground state.
There is, for example, the coherent de-excitation of
magneto-associated molecules \cite{winkler2007} in a STIRAP scheme
or the exploitation of electronic couplings in order to transfer
photoassociated molecules deeply into the singlet ground state
potential \cite{sage2005}. Recent milestones were the observation of
molecular condensates \cite{jochim2003} or the formation of
ultracold molecules in the v=0 internal ground state
\cite{sage2005,Lang2008,Ni2008,Deiglmayr2008}.

Although these methods have enjoyed recent success, it is still a
standing question as to how far coherent control of laser pulses can
be applied to form molecules and control their internal states. The
use of laser pulses, as opposed to cw light, has, in principle,
several advantageous properties.  A laser pulse has a large
bandwidth, enabling excitation of atom pairs to bound molecular
states resonantly over a large range of internuclear distances.
Calculations show that even simple chirping of such an excitation
pulse can have a large effect on the dynamics \cite{vala2001}.
Theoretically, a second short laser pulse (shorter than the
vibrational dynamics of the excited molecule) can take advantage of
the vibrational wavepacket dynamics resulting from such a broadband
excitation and efficiently stimulate the photoassociated molecules
into stable bound ground states \cite{koch2004}. Furthermore, by
(amplitude and/or phase) shaping both the excitation and
de-excitation pulses, the excited state wavepacket dynamics can be
altered to selectively populate ground state vibrational levels
\cite{poschinger2006} and increase overall efficiency using learning
algorithms \cite{judson1992}.

Although such schemes might offer many advantages theoretically, the
experimental implementation of them has proven to be more
complicated.  The use of chirped picosecond-length pulses for
photoassociation to excited states has shown some promise
\cite{fatemi2001} and a clear flux enhancement signal was observed.
However it is not clear how far coherent control can be applied to
such weakly bound wavepackets, since they theoretically have a
round-trip time of nanoseconds - comparable to the spontaneous decay
lifetime of the excited state (a decoherent process) and also
dephase rapidly due to potential anharmonicity. Chirped nanosecond
pulses have been investigated and found to alter ultracold atomic
collision rates \cite{wright2005:prl}. In the femtosecond regime
ultracold molecule dissociation and optimization of the dissociation
by shaped pulses has been experimentally verified
\cite{salzmann2006,brown2006}. These experiments highlight that
additional processes occur in the molecule-pulse interaction, which
counteract the PA process. Finding the right conditions under which
a photoassociation scheme becomes possible is the focus of ongoing
research and indeed also of this paper.

Other recent experiments have successfully used shaped femtosecond
pulses to control the photoassociative ionization of ultracold Rb
atoms in conjunction with cw fields \cite{veshapidze2007} and
multi-photon resonant ionization of ultracold ground state Rb$_2$
molecules \cite{weise2007}.  While molecular dynamics are involved
in both cases, neither of these experiments have shown the existence
of excited state photoassociated molecules which are suitable for
coherent stabilization. Coherent dynamics in the excitation of atoms
by ultrashort pulses was demonstrated in \cite{monmayrant2006},
which shows that electronic coherences can be created and detected
using ultrashort pulses. These may also be observable on vibrational
timescales (typically tens of picoseconds) when using ultrashort
pulses to photoassociate atoms.

In a recent paper \cite{salzmann2008} we showed that photoassociation with
shaped pulses is possible, and that coherent dynamics is visible.  The purpose
of the experiments was to investigate two points: The first was to answer the
question of whether it is possible to photoassociate atoms to bound excited
molecular states using shaped ultrashort pulses.  This process is the
bottle-neck in the transition from an atomic to molecular ultracold gas.  The
second was to observe and manipulate the coherent dynamics of the
electronically excited photoassociated molecules. With the dynamics known,
future experiments can be considered making use of this in conjunction with a
de-excitation pulse to stabilize the molecules.  In the current paper we give
a detailed description of the experiment and present the analysis of the
results of Ref. \cite{salzmann2008}. The comparison with the theoretical
description will be presented in our subsequent article \cite{merli2009}.

The paper is laid out as follows: In section \ref{sec:genconsid} we discuss
the general considerations one must be aware of before attempting a pulsed
photoassociation experiment on an ultracold sample.  In section
\ref{sec:expsetup} the experimental implementation is described. In section
\ref{ssec_interp} we discuss processes which compete with pulsed
photoassociation.  In section \ref{sec_fspa} the results of our experiments
are presented and we show that pulsed photoassociation has occurred with
coherent dynamics.

\section{General considerations}
\label{sec:genconsid}

Although many proposals have been published which discuss ways of using
ultrashort laser pulses for the efficient production of molecules from an
ultracold gas, the actual implementation is restricted by a number of
experimental limitations: The bandwidth and pulse shaping resolution of the
available laser system, the extra potential curves present in real molecular
systems (theoretically couplings between only two or three molecular
potentials are considered), the trapping and cooling process applied to the
atoms themselves and the influence of this on PA, and the detection of any
molecules produced.

Ideally, one would select a photoassociation pulse with suitable
bandwidth and tune its center frequency such that the total spectral
intensity of the pulse is distributed slightly below an excited
state potential asymptote, where the Franck-Condon factors (FCFs)
are large. Using picosecond laser pulses (as e.g. proposed by Koch
{\it et al.} \cite{koch2006}), which have a spectral bandwidth of
typically tens of wavenumbers, a suitable Condon point for
free-bound excitation of Rb lies at around 60 bohr radii.  Around
this internuclear distance FCFs are large enough to expect a
reasonable photoassociation efficiency. Unfortunately, for the
present experiments a tunable source of picosecond laser pulses was
not available.  Instead a femtosecond source was used, with a
correspondingly larger bandwidth.

In addition to photoassociating the atoms, we seek to manipulate the
free-bound excitation with optimally tailored laser pulses, as
proposed e.g. in \cite{poschinger2006}. Ultra-short pulse shaping
techniques are based on dispersive optics, such as double grating
pulse shapers, which, to date, are not available with suitable
resolution for narrow bandwidth picosecond pulses, and hence rules
out their use. However, the manipulation of femtosecond pulses is a
standard technique and offers a lot of freedom in controlling phase,
amplitude and polarization \cite{plewicki2007}.

For photoassociation experiments with femtosecond pulses, the large
bandwidth of several hundred wavenumbers requires further
consideration.  The restriction of spectral intensity to below the
potential asymptote is important in these experiments: Frequency
components on resonance with atomic D$1$ ($1257\,8$cm$^{-1}$) and
D$2$ ($12816\,$cm$^{-1}$) transitions cause intolerable losses of
atoms from the trap due to light scattering forces and ionization as
discussed in \cite{salzmann2006}. Frequency components blue to
atomic resonances can address anti-binding potential branches
leading to radiation shielding effects which are known to interfere
with cw photoassociation. All these effects lead to a strong
reduction in atomic density within the laser overlap, which will
inhibit successful photoassociation.

\begin{figure}
\begin{center}
\includegraphics[width=\columnwidth]{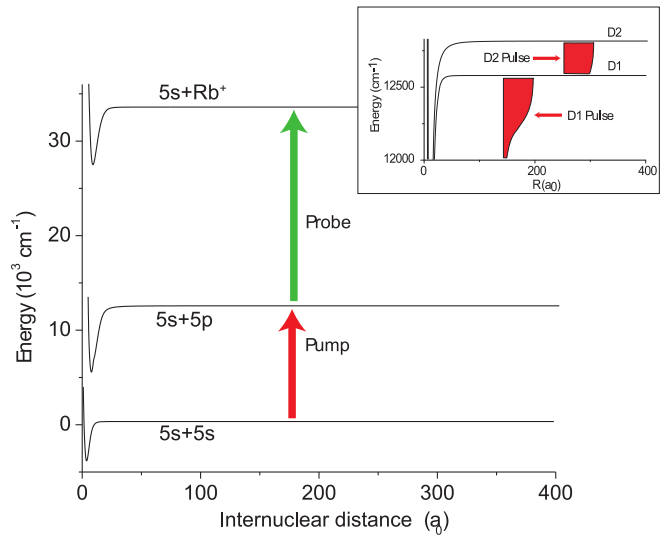}
\caption {\label{fig1_fsPAScheme} (color online) Scheme of pump-probe
  photoassociation. Formation of molecules in an excited state by exciting a
  pair of colliding ultracold atoms with a shaped pump-pulse. Transfer to
  molecular ion by a time-delayed probe-pulse.  Inset: Pulse spectra used when
  exciting to potentials with D1 (D1 pulse) and D2 (D2 pulse) asymptotes.}
\end{center}
\end{figure}

There are two ways to limit the spectral intensity at and above the
dissociation limit: Detune the central frequency of the pulse far
enough away from the dissociation limit such that the spectral
intensity above the dissociation limit is low enough \emph{or}
spectrally shape the pulse to eliminate these frequencies.  The
first method requires detuning the pulse on the order of
200\,\rm{cm}$^{-1}$ (for our available laser source). The
disadvantage of this method is that, although it does successfully
reduce the mentioned detrimental effects, the free-bound FCFs are
very low for such large detunings (by about two orders of magnitude
compared with those obtainable by using a suitable picosecond pulse
tuned closer to the dissociation limit).  In fact, the FCFs are only
reasonable as one approaches the dissociation limit, where the
spectral intensity of such a detuned pulse is unfortunately very
low, resulting in a poor overall photoassociation rate.  However,
the second method, spectrally shaping the pulse, results in high
spectral intensity close to the dissociation limit, where the
free-bound FCFs are large (how close one can get to the dissociation
limit depends on the shaping resolution), and therefore a good
photoassociation rate. This method also very effectively reduces the
mentioned detrimental processes. The photoassociation pulse central
frequency, therefore, need only be tuned a few tens of wavenumbers
below the molecular dissociation limit (the atomic D$1$ or D$2$
transition in our case). Spectral intensity at the potential
asymptote and above can be filtered out from the pulse spectrum by
an appropriate spectral amplitude transfer function, such as a step
or top-hat function with the sharp spectral cut-off closely below
the atomic resonance - i.e. a spectral low-pass or band-pass filter.
The inset of figure \ref{fig1_fsPAScheme} shows a schematic of
suitable resulting pulses for photoassociation below the D1 or D2
asymptote.

The next task is the detection of femtosecond photoassociated
molecules. The most reliable detection scheme is to ionize molecules
from the excited state directly after their formation by the pump
laser pulse. This allows the direct investigation of the molecule
formation with sub-picosecond resolution immediately after the
free-bound transition, thereby suppressing further interactions, for
example with trapping light. The ionization is done by a second
femtosecond probe pulse which is tuned to excite molecules from
their first excited to the molecular ionic state (see figure
\ref{fig1_fsPAScheme}). In such a pump-probe configuration,
photoassociation and ionization laser pulses are separated in time
by a defined, variable delay. Any ions produced can be mass-filtered
to remove the atomic ions produced by the lasers and thus
selectively view only molecular dynamics. For so-called pump-dump
experiments \cite{koch2006b, poschinger2006}, which aim to
coherently populate the molecular electronic ground state, this
pump-probe scheme may further give important information on nuclear
wave-packet dynamics in the excited state potential after
femtosecond photoassociation.

\section{Experimental procedure}
\label{sec:expsetup}

\begin{figure}
\begin{center}
\includegraphics[width=\columnwidth]{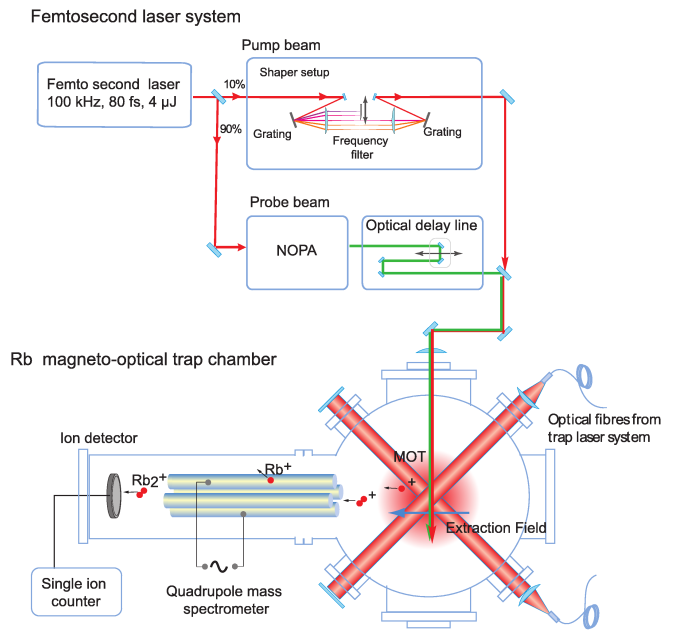}
\caption {(color online) Experimental setup for pump-probe photoassociation.
  Femtosecond pump pulses pass a zero dispersion grating pulse shaper,
  realizing an optical low pass to filter atomic resonances. Probe pulses are
  created by frequency conversion in a NOPA (non-collinear optical parametric
  amplifier) and pass a controllable optical delay stage. Pump and probe beams
  are spatially overlapped and focused into a rubidium dark-SPOT. Molecular
  ions created in the trap are mass selected by a radio frequency mass
  analyzer.  } \label{fig2_setup}
\end{center}
\end{figure}

The experiments were carried out in a collaboration project between the Freie
Universit\"at Berlin and the Universit\"at Freiburg. The joint experimental
setup (figure \ref{fig2_setup}) consists of a femtosecond laser system with a
pulse shaper, situated at the Institut f\"ur Experimentalphysik at the FU
Berlin, and a transportable high density magneto-optical trap for rubidium
atoms (a dark-SPOT).

The femtosecond laser pulses were produced by a Coherent Mira oscillator and
amplified by a Coherent RegA 9050, yielding pulses of 4$\,\mu J$ energy and an
autocorrelation of 80$\,$fs FWHM at 100 kHz repetition rate. Output pulses
have a spectral width of 390$\,$cm$^{-1}$ FWHM and are either centered around
12500$\,$cm$^{-1}$ (800$\,$nm), which is 78$\,$cm$^{-1}$ below the
dissociation limit of the first electronically excited 5s+5p$_{1/2}$ states,
or centered around 12739$\,$cm$^{-1}$ (785$\,$nm), which is 78$\,$cm$^{-1}$
below the dissociation limit of the electronically excited 5s+5p$_{3/2}$
states.

10\% of the laser output is split off and used for the pump pulses.  The pump
pulses pass through a zero dispersion, double grating pulse shaper equipped
with a spatial light modulator (CRI, SLM 640) to manipulate spectral phases
and amplitudes with a spectral resolution of 2.2$\,$cm$^{-1}$. Using the SLM
for attenuation of spectral components on resonance with rubidium transitions
proved to be insufficient as the 2\% residual transmission still causes
significant atom loss from the magneto-optical trap. The spectral low pass
filter was, instead, realized by placing an additional physical block in the
shaper's Fourier plane, spatially blocking the path taken through the shaper
by the high frequency part of the pulse spectrum. This block consists simply
of a razor edge (see figure \ref{fig2_setup}) which is mounted on a precision
stage. This setup provides a high-attenuation optical low-pass filter with a
sharp and adjustable cut-off frequency.

In order to adjust the cut-off of the filter to a suitable frequency we need
to know: First what the position versus frequency conversion of the precision
stage is, and, second, at what position reading of our high precision stage
the atomic resonance frequency is situated.  The answer to the former is given
by the zero dispersion shaper's properties (grating frequency and lens focal
length), whereas the answer to the latter is best measured experimentally.
The fluorescence of rubidium atoms in the magneto-optical dark spot trap was
measured on a photodiode as the cut-off was scanned.  The laser beam was
chopped (to minimize error due to atom-loss and trap fluctuations) and then
passed through the center of the cold cloud.  As soon as the spectral filter
transmits an atomic resonance frequency, the relative trap fluorescence (with
vs. without beam) increases. The resulting data has a step-like characteristic
at the resonance which is fitted by an error function to retrieve the
resonance position in the Fourier plane with a precision of about 86$\,\rm{\mu
  m}$, corresponding to a frequency resolution of 1.8$\,$cm$^{-1}$.

Ionization probe-pulses are produced by frequency conversion of the
RegA output in a non-collinear optical parametric amplifier (NOPA)
\cite{piel2006} which uses the remaining 90\% of the RegA output
power. The probe-pulses are centered at 20160$\,$cm$^{-1}$
(496$\,$nm) when using a D1 pulse and 19802$\,$cm$^{-1}$ (505 nm)
when using a D2 pulse, have a spectral FWHM of 1000$\,$cm$^{-1}$ (25
nm), pulse energies of up to 50 nJ and non-transform-limited
autocorrelation durations of about 600$\,$fs FWHM. To vary the
pump-probe delay the probe pulses pass an optical delay stage
(PI,M-531). During the pump-probe scans the delay is typically
scanned with a speed equivalent to a delay change of 15 fs/s.

Before entering the vacuum chamber of the magneto-optical trap, both
beams are spatially overlapped using a dichroic mirror and focused
into the trap to waists of $\approx$120 $\mu$m and $\approx$100
$\mu$m, resulting in typical peak intensities of
$2.5\times10^4$\,MW/cm$^2$ for the pump and $1.4\times
10^3$\,MW/cm$^2$ for the probe pulses.

The magneto-optical dark SPOT \cite{ketterle1993,townsend1996,anderson1994}
captures $10^8$ $^{85}$Rb atoms at densities of $10^{11}\,$cm$^{-3}$ and
temperatures of 100$\,\mu$K. Trap densities and sizes are measured by
absorption imaging, the trap fluorescence is continuously monitored by a
photodiode.  Trapping light is produced by two single-mode diode lasers which
are actively stabilized to frequency modulation spectroscopy signals
\cite{bjorklund1983}. The trapping laser is detuned by 3$\,\Gamma$ (18 MHz)
from the F=3$\rightarrow$F'=4 transition and reaches a peak intensity of
18mW/cm$^2$. The second laser is stabilized to the F=2$\rightarrow$F'=3
transition for repumping.

The key parameter of a dark SPOT is the steady state ratio of upper
to lower hyperfine ground state population of the trapped atoms
($p=N_{\rm res}/N_{\rm total}$), where $N_{\rm res}$ is the number
of atoms in a state near resonance with the trapping light and
$N_{\rm total}$ is the total number of atoms. In $^{85}$Rb the
relevant levels are the F=2 and F=3 hyperfine ground states, which
are split by 3.1$\,$GHz.  For the dark SPOT configuration, repump
laser intensity is removed at the position of the trap by means of
two crossed, hollow beams of repump light. This allows optical
pumping of population from the upper to the lower hyperfine ground
state via off-resonant excitation from the F=3 electronic ground
state to the F'=3 electronically excited state by the trapping laser
in this region.  The effect of residual repump light in the hollow
beam overlap is reduced by detuning the laser by 30$\,$MHz red to
the F=2$\rightarrow$F'=3 transition. Atoms can be optically pumped
between the hyperfine states by two additional laser beams which
fill the trap center only. A depumper on resonance with the
F=3$\rightarrow$F'=2 transition decreases the F=3 population and a
separate beam of repump light increases it. In order to measure the
relative populations in both hyperfine states, all trapped atoms are
pumped to the upper F=3 state by a 50$\,$ms pulse of the repump fill
beam, resulting in a flash of fluorescence which is detected by the
photodiode \cite{townsend1996}. The population ratio $p$ is then
deduced from the fluorescence signals in the dark- and repumped
trap, assuming that the trap fluorescence is proportional to the F=3
population and the fluorescence is dominated by decay from the F'=4
excited state. In our setup, the detuning of the repumper by
30$\,$MHz alone causes 90\% of the atoms to be kept in the lower F=2
hyperfine state (without the additional depumper beam), which
provides the highest densities and therefore optimal conditions for
photoassociation experiments. In our experiments we vary the
parameter $p$ to exclude competing processes such as
photoassociation by trapping light (see sec. \ref{ssec_interp}).

Atomic and molecular ions produced in the trap are extracted by an
electric field of 40$\,$V$/$cm and detected by a channeltron. Before
the channeltron a radio frequency (rf) quadrupole filters the
incoming ions according to their mass (see figure \ref{fig2_setup}).
Alternatively we can operate the mass selection as a time-of-flight
spectrometer. Electronic ion signals are high pass filtered to
remove any residual rf signal from the mass filter before they are
amplified and digitized by a constant fraction discriminator.
Digital pulses are acquired by a fast counter and integrated over
100\,ms. The dark count rate (without lasers and trapped atoms) of
the ion detection system is 0.1 Hz. During pump-probe scans the ion
count rate, trap fluorescence and the actual pump-probe delay are
gathered by the data acquisition computer.

Despite blocking the atomic resonances, the femtosecond pulses produce large
numbers of atomic ions by off-resonant pump-probe excitation. Without mass
filtering by the rf quadrupole, these would totally saturate the detection
system and prohibit the identification of small numbers of molecular
ions. Mass spectra of the pump-probe created ions in the trap, using the rf
quadrupole, show two peaks at the atomic $^{85}$Rb and the molecular
$^{85}$Rb$_2$ masses with resolutions of 4 amu. For different mass settings
only the system's dark count rate is detected. This shows that ion signals
measured on the $^{85}$Rb$_2$ mass represent exclusively rubidium molecular
ions and no background due to different charge states or ion species has to be
considered.

\section{Competing processes}

\label{ssec_interp} The identification of femtosecond
photoassociation relies very strongly on understanding alternative sources of
the molecular ion signals detected during the experiments.  Two potential
alternative pathways exist for producing Rb$_2^+$ ions - a) excitation of
trap-light-produced molecules and b) collisional autoionization via Rydberg
states.

Photoassociation of rubidium dimers can occur due to trap light
\cite{marcassa2005}. These molecules may be ionized by the
femtosecond pulses from the electronic ground- (the pump pulse + the
probe pulse) or excited- (the probe pulse only) state. In a dark
SPOT the efficiency of trap light photoassociation depends on the
atomic population of hyperfine ground states. This distribution can
be characterized by the ratio, $p$, of atoms in the upper F=3 state,
which participate in the trapping cycle, to the total number of
atoms. The density of atoms varies over about one order of magnitude
as $p$ changes and reaches its maximum around $p=0.1$
\cite{townsend1996}. As $p$ is varied, the photoassociation rate for
both femtosecond and trap light PA changes due to its dependence on
the density squared. Because of the tight bandwidth of the trapping
lasers, photoassociation by trap light is additionally sensitive to
the hyperfine configuration of the colliding pairs \cite{helm2004},
which is directly influenced by $p$, whereas the broadband
femtosecond pulses are not. The examination of molecular ion rates
under variation of the atomic hyperfine ground state populations is
a good measure to distinguish trap-light-formed and
femtosecond-formed molecules.

\begin{figure}
\begin{center}
\includegraphics[width=\columnwidth]{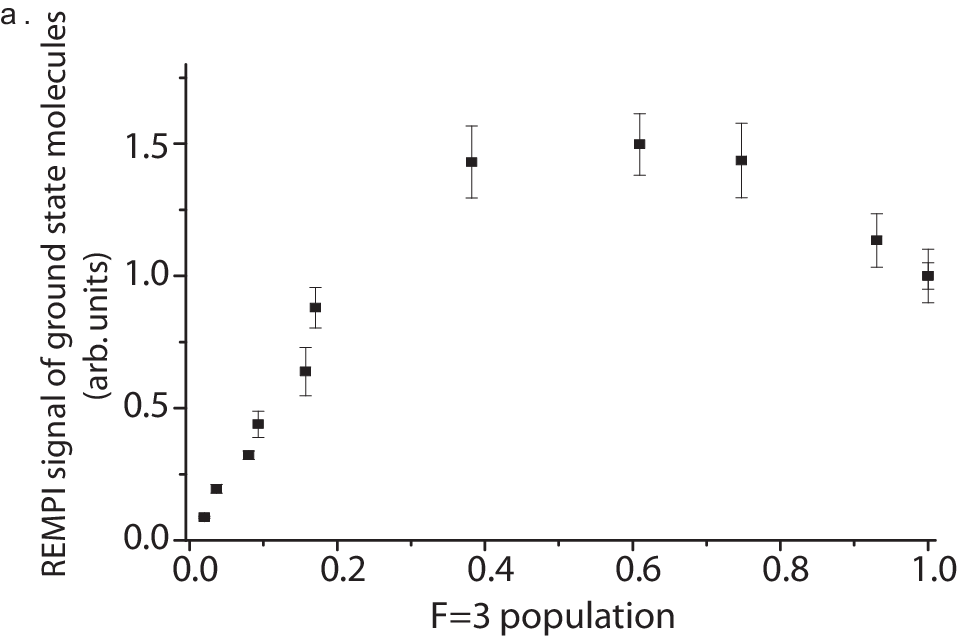}
\includegraphics[width=\columnwidth]{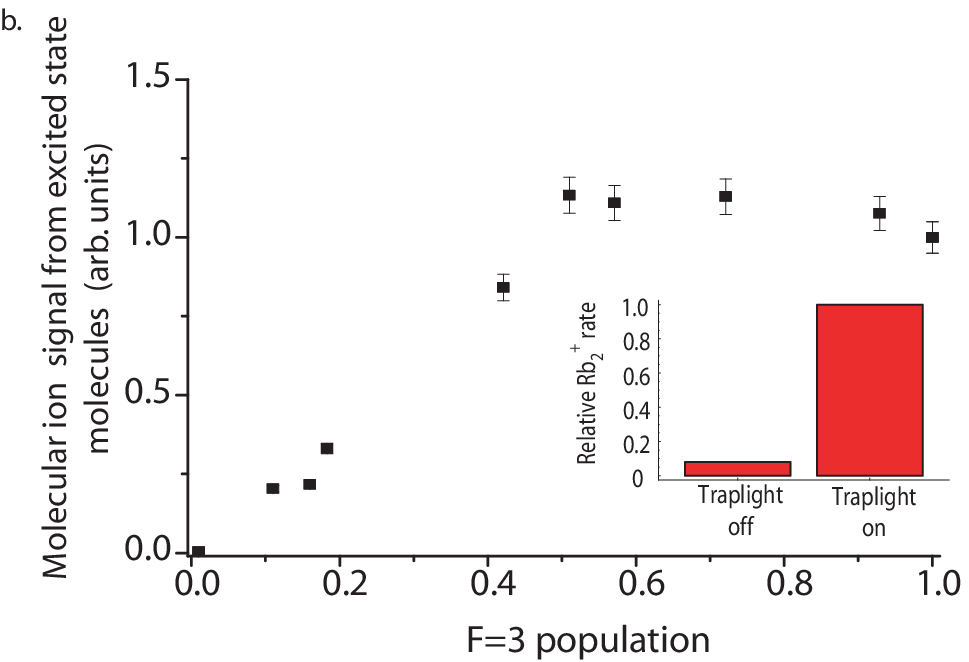}
\caption {a. Molecules photoassociated by trapping light, ionized from 5s+5s
  ground state by nanosecond dye laser pulses. b.  Molecules produced by
  trapping light, ionized from the 5s+5p$_{3/2}$ excited state by femtosecond
  probe pulses.  Inset: Relative Rb$_2^+$ rate with and without trapping light
  present simultaneously with the femtosecond pulses for
  $p=0.01$.} \label{fig3_Dye_vs_BSF}
\end{center}
\end{figure}

The trapping laser photoassociates atom pairs to states below the
5s+5p$_{3/2}$ potential asymptote. After a lifetime of 12\,ns they
spontaneously decay to their electronic ground state, thereby
populating the uppermost vibrational levels of the $^3\Sigma_u^+$
5s+5s metastable state \cite{marcassa2005,gabbanini2000}.  The
contributions of the excited state trap-light-formed molecules (1
probe photon) and the ground state trap-light-formed molecules (1
pump + 1 probe photon) to the molecular ion signal are investigated
separately.

For an independent measurement of the dependence of
trap-light-formed ground state molecules on the F=3 population, we
use resonant two-photon excitation (REMPI) for state selective
detection \cite{gabbanini2000}. A Nd:YAG pumped dye laser, operated
at 602$\,$nm with 10mJ pulse energy and a 10Hz repetition rate, is
focused into the trap. The detected molecular ion rates were on the
order of 0.5 per pulse for a trap with 100\% F=3 population ($p=1$).
The formation of ground state molecules as a function of $p$ is
shown in figure \ref{fig3_Dye_vs_BSF}a. The molecular signal is low
at small p, reaches its maximum at about $p=0.6$ and drops again for
$p=1$.  The trap density has a very different $p$ dependence, and
peaks around $p=0.1$ \cite{townsend1996}. Thus the comparatively
slow rise in figure \ref{fig3_Dye_vs_BSF}a to $p=0.6$ can be
attributed to a photoassociation process that requires at least one
of the colliding atoms to be in the F=3 hyperfine state.

A similar $p$ dependence is observed for the molecular ion signal which is
detected when only the femtosecond probe pulses interact with the trap (figure
\ref{fig3_Dye_vs_BSF}b). This signal rises to reach a maximum at around
$p=0.5$ and drops again as $p$ approaches unity. The rate of molecular ions
which are detected this way is up to 500\,Hz.

By shuttering both the trapping light and the femtosecond probe beam
it is possible to determine the origin of these ions.  In one case
the femtosecond probe-pulses were incident on the MOT only when the
trap light was shuttered off (the trapping light had a large on/off
duty cycle in order to avoid density loss).  In the other case, the
timing of the shuttering of the femtosecond probe-pulses was shifted
so that they were incident on the MOT when the trap light was on
(the trap light had the same duty cycle as the first case in order
to have comparable conditions). The ion signal is strongly
correlated to the presence of trapping light, as shown in the inset
of figure \ref{fig3_Dye_vs_BSF}b for the two shuttering schemes.  If
the trapping lasers are off when the probe-pulses hit the trap, the
Rb$_2^+$ signal is only around 10\% as large as if the trapping
lasers are on when the probe-pulses hit the trap.  The reason for
this is that as soon as the trapping lasers are shuttered off, no
more molecules can be photoassociated to the excited state and the
excited state molecules present at that time decay with a
spontaneous lifetime of 12ns. After this decay time there are no
molecules in the excited state until the trap light is shuttered on
again. The strong reduction in molecular signal observed when the
molecules are in the ground state shows that the probe pulses can
ionize molecules from the excited 5s+5p state, but not from the
ground state.  We therefore attribute the signal from figure
\ref{fig3_Dye_vs_BSF}b to molecules which are photoassociated by
trapping light into the 5s+5p$_{3/2}$ potential and which are then
ionized by the femtosecond probe pulses.  This explains the
similarity of the $p$ dependence found in figures
\ref{fig3_Dye_vs_BSF} a and b, as the 5s+5p$_{3/2}$ population
represents the intermediate step in the formation process of ground
state molecules by trapping light. In the pump-probe experiments the
trapping light was \emph{not} shuttered, and therefore there was a
steady-state population of trap-light-photoassociated molecules,
which are also ionized by the probe laser and detected, causing a
background molecular ion signal. This background is measured
separately and may be corrected for.

The second process that can lead to the formation of molecular ions is the
associative autoionization in collisions of Rydberg and ground state atoms
\cite{barbier1987}. Rydberg atoms are created from ground state atoms by
excitation with a pump and a probe pulse.  The resulting distribution of
Rydberg states peaks at principal quantum number of n=12. Collisional
associative ionization (CAI) may occur during the spontaneous lifetime of the
Rydberg state, following the reaction: Rb(nl)+ Rb(5s)$\rightarrow$ Rb$_2$ +
e$^-$.  The contribution of this process to the observed molecular ion rate
can be estimated from a capture model \cite{levine2005}. The required C$_6$
coefficients were calculated following \cite{marinescu1997}, where the
necessary ground state wave functions were obtained using a model potential
described in \cite{marinescu1994}. In combination with the trap temperature of
100\,$\mu$K, we estimate a rate coefficient for associative ionization of
k$\,=\,4\cdot10^{-10}$\,cm$^3/$s.  An upper limit on the CAI rate can be
estimated by attributing the total atomic trap loss rate at positive
pump-probe delays to the excitation to Rydberg states. In reality, trap-loss
involves many processes (for example atomic ionization) and so the actual rate
is most certainly lower than this. With the measured atomic density of
$10^{11}$ cm$^{-3}$ and an average lifetime of the addressed Rydberg levels of
1.5\,$\mu$s \cite{gallagherbook1994} we estimate the maximum contribution of
this process to be less than 20\% of the molecular ion signal detected in the
pump-probe experiments. Furthermore, this process can only occur at positive
delays (pump before probe) when the first step in the Rydberg excitation
proceeds via the pump pulse. At negative delays, where significant numbers of
molecular ions are also detected (see section \ref{ssec_negdel}), the Rydberg
excitation would have to proceed via an initial excitation by the probe
pulses, however no appropriate atomic transitions are available for such a
process.

\section{Femtosecond photoassociation}
\label{sec_fspa}

\subsection{Pump-probe spectra}
\label{ssec_ppspec}

\begin{figure}
\begin{center}
\includegraphics[width=\columnwidth]{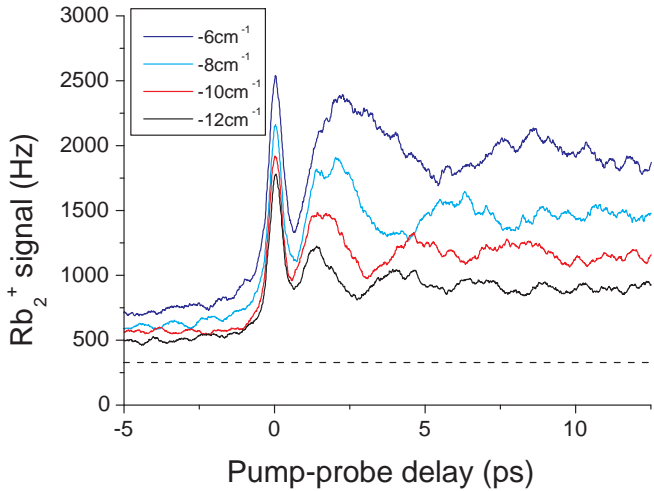}
\caption{(color online) Dynamics of Rb$_2^+$ signal with a D1
excitation pulse. Pump-probe
  data for cut-off detunings of -6 cm$^{-1}$ to -12 cm$^{-1}$ from the D1
  asymptote.  The dashed line indicates the background level (see
  text).} \label{fig6_ppdata_sim}
\end{center}
\end{figure}

Figure \ref{fig6_ppdata_sim} shows measured molecular ion count
rates as a function of the delay between a pump-pulse tuned below
the D1 asymptote (see ``D1 pulse" inset figure
\ref{fig1_fsPAScheme}) and an ionization probe-pulse. Going from top
to bottom, the spectral cut-off frequency, relative to the D1
transition frequency (the cut-off detuning), decreases from
-12\,cm$^{-1}$ to -6\,cm$^{-1}$ (the minus sign means red to the D1
transition) in steps of 2\,cm$^{-1}$.  The data were smoothed by
averaging over 10 adjacent points. The dashed lines indicate an
estimate of the background ion rate caused by probe laser ionization
of trap-light photoassociated molecules (see section
\ref{ssec_interp}).  In this particular set of measurements the
background level was not measured, rather estimated by extrapolating
the power-dependence of the ratio of the negative-delay level to the
measured background level of a previously-measured data set.

The data show that the detected molecular ion rate clearly depends
on both the pulse delay and the pump pulse's spectral cut-off
frequency. The general form of the curves is the following: For
negative delays (when the ionization probe-pulse precedes the pump
pulse) we observe a constant rate of molecular ions. At t=0 both
pulses coincide in time and the count rate increases drastically,
forming a peak of 0.5\,ps width. For positive delays, a clear
increase in the molecular ion signal, compared to negative delays,
is observed. Additionally, the signal is modulated by characteristic
oscillations with periods of a few picoseconds, which are fully
damped after a short time. As the cut-off detuning is decreased
(going from top to bottom in figure \ref{fig6_ppdata_sim}) the
molecular ion signal also increases for all delays, as does the
modulation period for t$>$0. Significant pump-probe signals are only
observed in the experiments for cut-off detunings of less than -30
cm$^{-1}$.

\begin{figure}
\begin{center}
\includegraphics[width=\columnwidth]{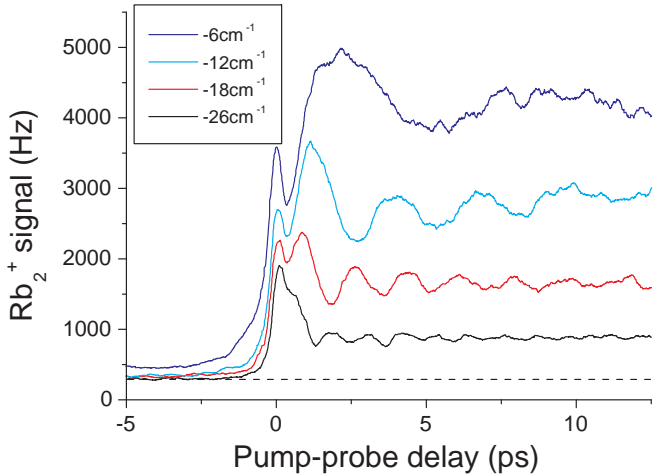}
\caption{(color online) Dynamics of Rb$_2^+$ signal with D2
excitation pulse. Pump-probe
  data for cut-off detunings of -6 cm$^{-1}$ to -26 cm$^{-1}$ from the D2
  asymptote.  The dashed line indicates the background level (see
  text)} \label{fig6_ppdata_simD2}
\end{center}
\end{figure}

Figure \ref{fig6_ppdata_simD2} shows the corresponding results with
the ``D2 pump-pulse" (see inset figure \ref{fig1_fsPAScheme}) and an
ionization probe-pulse. Except for the overall count rate,
quantitatively identical data are obtained on the D$2$ asymptote as
are obtained on the D1 asymptote. For experiments on the D$2$
asymptote, the frequencies red to and including the D$1$ resonance
frequency were additionally blocked as shown in the inset of figure
\ref{fig1_fsPAScheme}.  This second cut-off frequency was typically
$20\,\rm{cm}^{-1}$ blue to the D1 atomic resonance. Interestingly,
this suggests that the dynamics of our process is not sensitively
dependent on the form of the addressed potential.  The background
level was estimated to be similar to the negative-delay level for
the $-26\,\rm{cm}^{-1}$ data set.

With our setup one might expect to observe the propagation of a nuclear
wave-packet (which is created by the pump pulse and subsequently oscillates in
the excited state potential well) under the assumptions that a) a
well-localized wave-packet is created, b) the dephasing of the wave-packet is
not rapid and c) one looks on the appropriate time-scale.  The ionization
pulse would result in a delay-dependent molecular ion signal due to motion on
the internuclear degree of freedom.  Based on the spectral components of our
pump-pulse and with knowledge of all the potentials corresponding to this
asymptote, we expect the nuclear oscillation time to be on the order of tens
of picoseconds, rather than the few picoseconds we observe. Even pump-probe
scans over longer delay ranges do not show oscillations on the expected
time-scales. Since the observed oscillations are therefore not consistent with
a vibrational wave-packet, other processes must be considered in order to
explain the observed signals.

To get quantitative information for further analysis we extract characteristic
parameters from the data. In order to obtain the mean value of molecular ion
counts for negative delays, we take an average of data points for
t$<$-3\,ps. The modulation periods at positive delays are extracted by fitting
a damped oscillatory function to the data, with a steady-state asymptotic
level, a modulation amplitude, a damping time and an oscillation period and
phase relative to t=0. An additional linear slope parameter accounts for a
slow linear variation in the detection rate caused by a slow systematic drift
in the apparatus during the scan.

\begin{figure}[h]
\begin{center}
\includegraphics[width=\columnwidth,angle=0]{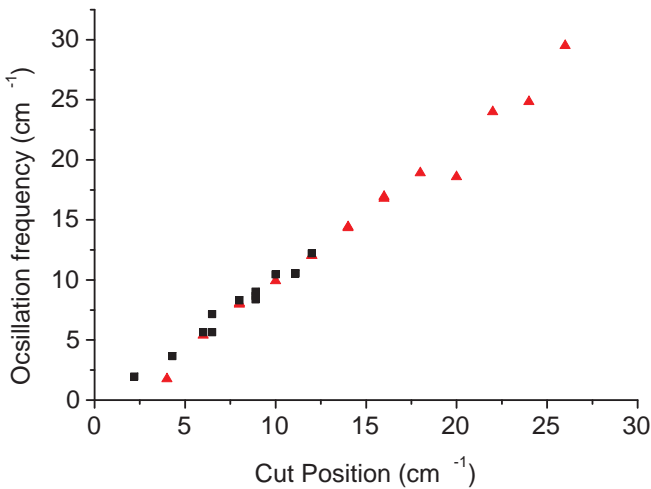}
\caption{(color online) Frequencies of modulations on pump-probe ion signals
  at $t>0$ versus cutoff position for both D1 (black squares) and D2 (red
  triangles) pulses.} \label{fig7_Feq_cutoff}
\end{center}
\end{figure}

A correlation plot (see figure \ref{fig7_Feq_cutoff}) of the fitted modulation
frequencies shows that the modulation frequency matches the cut-off detuning.
This behavior is not expected for a vibrational wave-packet oscillating in an
anharmonic molecular potential.  The observed oscillatory dynamics are instead
due to so-called coherent transients \cite{monmayrant2006,salzmann2008}, which
is a coherent energy exchange between molecules and the pump-pulse laser
field. A detailed discussion of this is given in our accompanying theoretical
paper \cite{merli2009}.

\subsection{Asymptotic behavior}
\label{ssec_ppasym}

\begin{figure}
\begin{center}
\includegraphics[width=\columnwidth]{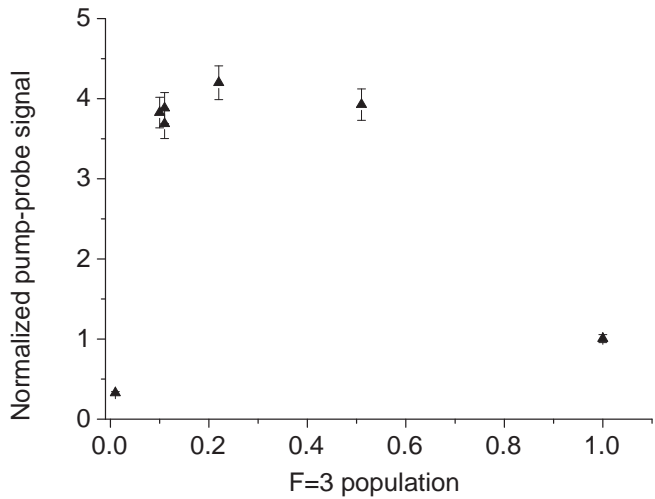}
\caption{(color online) Pump-probe mean molecular ion signal at
$t\gg0$ as function
  of dark SPOT F=3 population for D1 pulse.  A similar curve is obtained for
  the D2 pulse.} \label{fig8_PP_vs_BSF}
\end{center}
\end{figure}

In order to confirm that the pump-probe molecular ion signal at positive
delays originates from the excitation of free atom pairs by the femtosecond
laser and not from trap-light photoassociated ground state molecules, we
performed pump-probe scans at different configurations of the dark SPOT -
i.e. different values of the parameter $p$. These measurements can be compared
to those found in section \ref{ssec_interp} (figure \ref{fig3_Dye_vs_BSF}).

Figure \ref{fig8_PP_vs_BSF} shows the dependency of the asymptotic
molecular ion signals ($t \gg 0$) as a function of $p$. The signal
shows a steep increase from $p=0$, reaches its maximum at about
$p=0.2$ and drops for $p$ approaching unity. This curve looks very
similar to the atomic density curve in a dark SPOT
\cite{townsend1996} and so the pump-probe signals only seem to be
sensitive to the change in density as $p$ rises and not the
population in the F=3 state - consistent with photoassociation from
the femtosecond laser. This can be contrasted with the measurements
shown in figure \ref{fig3_Dye_vs_BSF}, showing photoassociation by
trap light, where the much slower rise for small $p$ indicates a
process which requires at least one of the colliding atoms to be in
the F=3 state.

We therefore conclude that the molecular ion signals from the pump-probe scans
must predominantly originate from the photoassociation of colliding atom pairs
by the femtosecond pump pulses.  In light of this, the pump-probe data
presented in this work were taken at $p=0.1$ where we measured highest trap
densities and trap light photoassociation rates are low, according to figure
\ref{fig3_Dye_vs_BSF} a, so that their contribution is, if present, of minor
importance.

\begin{figure}
\begin{center}
\includegraphics[width=\columnwidth]{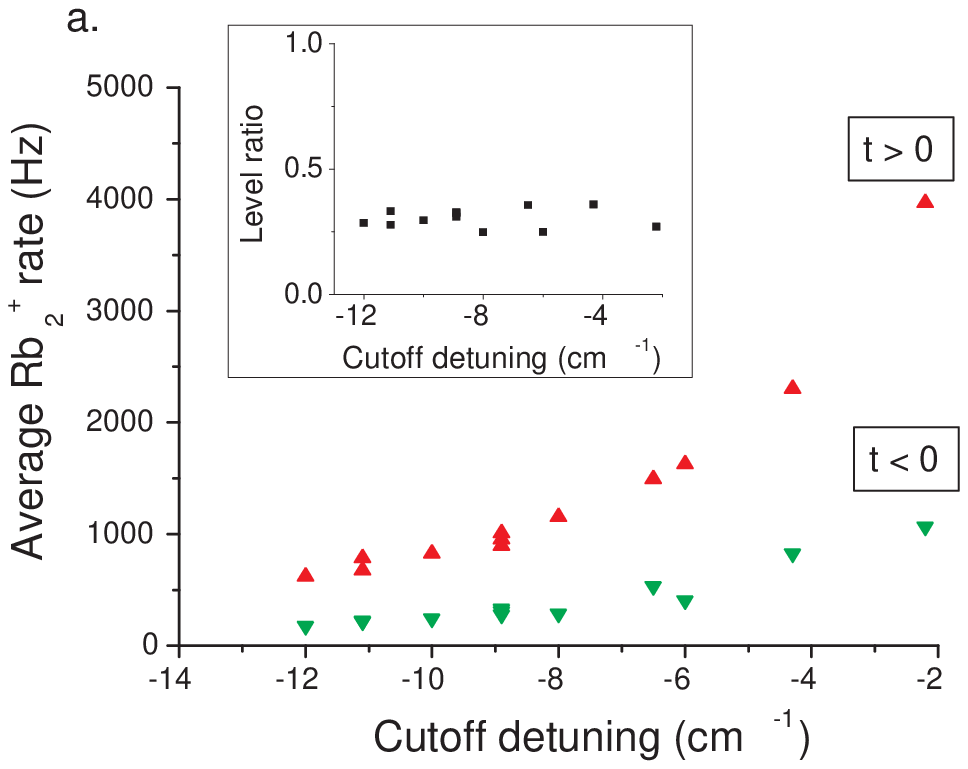}
\includegraphics[width=\columnwidth]{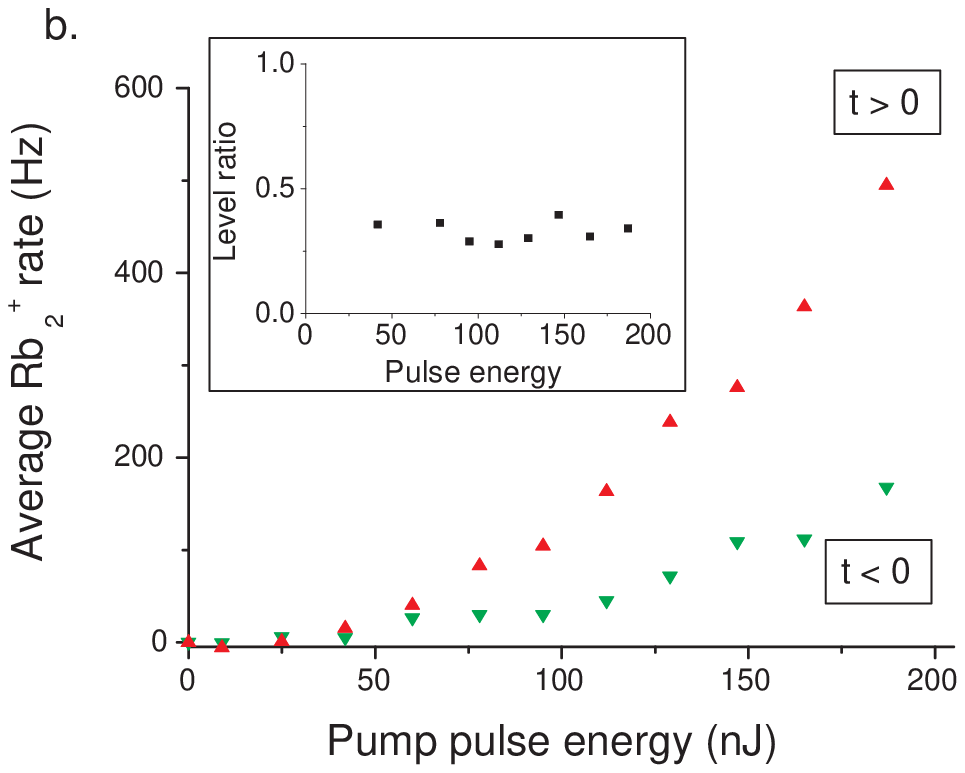}
\caption {a.) D1 pump-probe Rb$_2^+$ asymptotic rates versus
spectral cutoff
  position. $\blacktriangle$: Rb$_2^+$ at positive delays.
  $\blacktriangledown$: Rb$_2^+$ at negative delays. Inset: Constant ratio of
  $\sim0.3$ between time-averaged levels at negative and positive delays for
  different cutoff detunings. b.) Pump-probe Rb$_2^+$ average rates pump-pulse
  energy. $\blacktriangle$: Rb$_2^+$ at positive delays. $\blacktriangledown$:
  Rb$_2^+$ at negative delays. Inset: Constant ratio of $\sim0.3$ between
  time-averaged levels at negative and positive delays for different
  pump-pulse energies} \label{fig9_PP_levels}
\end{center}
\end{figure}

Figure \ref{fig9_PP_levels}a. shows the variation of the asymptotic pump-probe
signals with the spectral cut-off detuning. A clear increase in the Rb$_2^+$
rate is observed as the cut-off approaches the potential asymptote. As
discussed previously, the signal at t$>$0 represents the population in the
5s+5p manifold. For smaller cut-off detunings, the coupling of the electronic
states by the pump pulse increases, resulting in a higher excitation
efficiency. In the pump-probe scans, since the excited state population is
larger, the detected ion rate will be correspondingly higher.

Figure \ref{fig9_PP_levels}b. shows the variation of the asymptotic
Rb$_2^+$ rates as a function of pump-pulse energy. The signals rise
non-linearly with the pump-pulse energy.  Although only the
asymptotic values ($\Delta t \gg 0$) have been shown, the signal
gets larger for all delays, as the pump-pulse energy is increased.
This power dependence indicates a non-linear character of the pulsed
PA process.

\begin{figure}
\begin{center}
\includegraphics[width=\columnwidth]{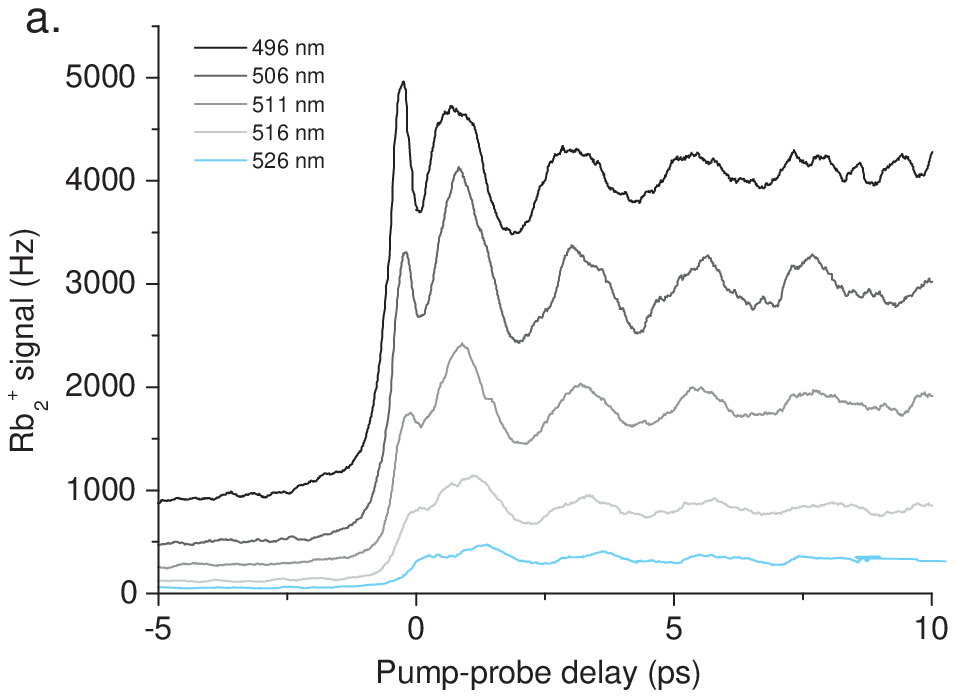}
\includegraphics[width=\columnwidth]{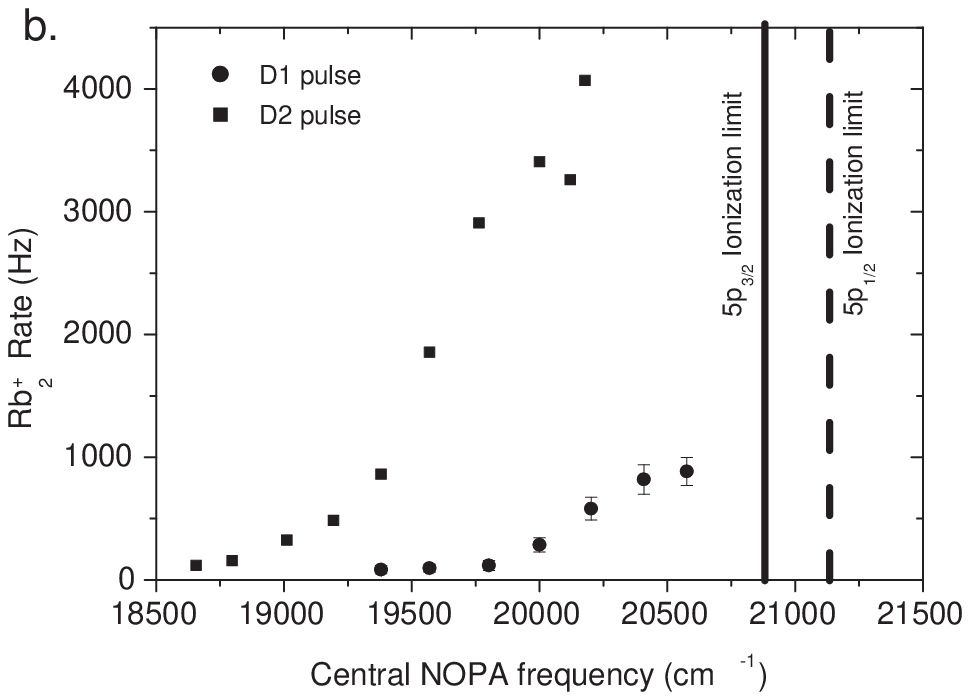}
\caption {a) Pump-probe scans for different center values of the NOPA
  (ionization) pulse using a D2 pulse. b) Rb$_2^+$ signal at $t\gg0$ for
  different center frequencies of the NOPA pulse.}
\label{fig5a_NOPA_wavelength}
\end{center}
\end{figure}

Figure \ref{fig5a_NOPA_wavelength}a. shows the result of pump-probe
scans using a D2 pulse with a cut-off of -14cm$^{-1}$ as the NOPA
center frequency is changed.  The temporal broadening for red
detuned probe pulses, particularly visible for the peak at zero
delay, can be explained by the increased duration of the probe-pulse
for higher center wavelengths.  Figure \ref{fig5a_NOPA_wavelength}b
shows the steady-state levels at $t\gg0$ using D1 pump-pulse and a
D2 pump-pulse as the NOPA center frequency is changed. Clearly, the
molecular ion rate drops as the center frequency is moved further
away from the dissociation limit (vertical lines) for both the D1
and D2 pulses, which supports the 1+1 REMPI scheme. With a D2 pulse,
the signal begins to increase at a NOPA central frequency of around
$18750\,\rm{cm}^{-1}$, $2124\,\rm{cm}^{-1}$ away from the
dissociation limit, which is $20874\,\rm{cm}^{-1}$ above the
5p$_{3/2}$ level.  With a D1 pulse, the signal begins to increase at
a NOPA central frequency of around $19800\,\rm{cm}^{-1}$,
$1312\,\rm{cm}^{-1}$ away from the dissociation limit, which is
$21112\,\rm{cm}^{-1}$ above the 5p$_{1/2}$ level. (The dissociation
limit is calculated as the atomic ionization limit of
$33690.81\,\rm{cm}^{-1}$ minus the asymptotic energy of the
populated molecular states - i.e. $12578.950\,\rm{cm}^{-1}$ for the
5p$_{1/2}$ states and $12816.545\,\rm{cm}^{-1}$ for the 5p$_{3/2}$
states, all relative to the atomic ground 5s$_{1/2}$ state energy.)
The depth of the ionic potential is approximately
6000\,\rm{cm}$^{-1}$ \cite{aymar2003}, so, in principle, it would be
possible to populate more deeply bound levels in the ionic potential
with the available laser frequencies. From our data it seems that
this is not the case, rather we populate mid- to weakly-bound states
in the molecular ion potential.  After the photoassociation pulse we
expect, due to laser spectrum and FCF considerations, to be mostly
populating states within approximately a few tens of wavenumbers of
the 5p$_{1/2}$ or 5p$_{3/2}$ dissociation limit (in fact, as will be
shown in our follow-on paper, we are mainly populating levels even
closer to the dissociation limit of the 5s+5p states.) Such weakly
bound molecules will be excited most efficiently to reasonably
weakly bound ionic states.

\subsection{Signal at negative delays}
\label{ssec_negdel}

Molecular ion levels at $t \gg 0$ and $t \ll 0$ show highly correlated
behavior over a broad range of cut-off detunings and pump-pulse energies. The
insets in figures \ref{fig9_PP_levels} a and b show the ratios of asymptotic
levels of molecular ion rates for $t \gg 0$ and $t \ll 0$ as the cut-off
detuning and the pump-pulse energy is varied. This ratio is fairly constant at
$0.3$ for both cases.  It is therefore possible that the ions detected at
positive delays and negative delays have a similar origin, with just a
different overall ionization efficiency.  PA due to the probe-pulses is not
expected to take place, due to the lack of a suitable transition for ground
state pairs and can be therefore ruled out. The signals are clearly different
in that no oscillations are seen for negative delays, which suggests a
possible mechanism which is consistent with the observed behavior.  The
molecular ions which are detected at negative delays must be first
photoassociated into the 5s+5p state by the pump-pulse before undergoing
further transitions and ending up as molecular ions.  At negative delays the
probe- precedes the pump-pulse, so any signal related to a 5s+5p excitation
must be due to a previous pair of pulses. The temporal separation between
pulse pairs of 10\,$\mu s$ is much longer than the 12\,ns spontaneous lifetime
of the 5s+5p state. Therefore, any population in the first excited state,
created by a pump pulse, will have decayed before a probe pulse of a
successive pair of pulses interacts with the trap.

The ionization of residual population in the excited state can therefore be
excluded as the origin of the negative delay signal.  However, the negative
delay signal can be attributed to molecules in the electronic ground
state. These are formed by spontaneous decay from bound excited states,
previously photoassociated by a pump pulse. The pump-pulse excitation results
in a considerable population in bound vibrational states. Upon decay these may
either dissociate or end up in bound molecular states. Vibrational ground
states are populated according to their Franck-Condon overlaps with the 5s+5p
vibrational states. As the 5s+5p molecules are loosely bound and of long range
character, ground state molecules can also be expected to populate the
uppermost vibrational states in the ground state potential. In
cw-photoassociation experiments \cite{gabbanini2000,lozeille2006} this process
is well known and the major formation process of ultracold molecules in their
electronic ground state. In our experiments, ground state molecules are
detected by excitation to an intermediate state below the 5s+4d or 5s+6s
asymptotes by a probe-pulse and subsequently ionized by a pump-pulse. The
ratio of 0.3 between the asymptotic levels at positive and negative delays is
the relative efficiency for direct pump-probe ionization versus ionization via
formation of ground state molecules.  The lack of oscillations on this signal
is due to the long time between formation and ionization with an incoherent
spontaneous emission step in between.

\subsection{Influence of linear chirp}

\begin{figure}
    \centering
    \includegraphics[width=\columnwidth]{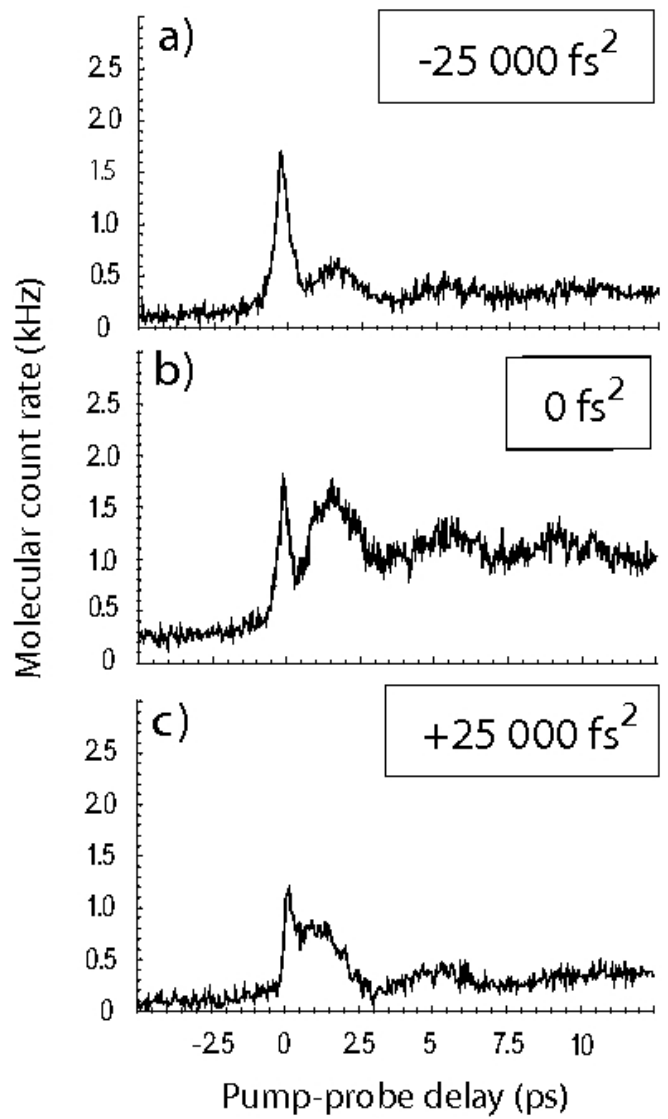}
    \caption[Cut pulse chirp]{Pump-probe Rb$_2^+$ rate for chirped and cut pulses with spectral cut-off at -8 cm$^{-1}$. The linear chirps are: a) -25000 fs$^2$, b) no chirp and c) +25000 fs$^2$.}
    \label{fig12_Chirped_pulses}
\end{figure}

A topic of great interest in photoassociation by femtosecond pulses
is the proposed possibility to coherently control the process by
imprinting defined phases on the molecular quantum state to
manipulate its temporal evolution. The application of linear chirps,
for example, is expected to increase the excitation efficiency of
the free bound transition in the perturbative regime
\cite{vala2001,luc-koenig2004,koch2006c,wright2007,e_shapiro2007} by
adiabatic transfer. Also the nuclear dynamics can be influenced by
imprinting the field's phase onto a coherent vibrational wavepacket
in the excitation. As demonstrated by Koch {\it et al.}
\cite{koch2006b} or Poschinger {\it et al.} \cite{poschinger2006}
this may allow optimization of the coherent formation of bound
ground states in pump-dump experiments.

In our pump-probe photoassociation experiments, we are not in the perturbative
regime, as can be seen from our non-linear power dependence in figure
\ref{fig9_PP_levels}.  Additionally our signal does not originate from nuclear
wave-packet motion. Nevertheless, chirping the pump-pulses reveals interesting
dynamics. We modified the pump pulse's phase by applying a quadratic spectral
phase (using the SLM in the Fourier plane in addition to the knife-edge) in
order to investigate the effect on the pump-probe signal.

The spectral cut-off, in combination with the quadratic phase,
causes strong distortions of the pulse's amplitude and frequency
behavior during the pulse maximum.  Compared to an unchirped
pump-pulse, the maximum field strength is reduced by a factor of
$\sim$2 and its duration is increased to 1.2\,ps FWHM. The
instantaneous frequency varies strongly on a sub-picosecond
timescale and only the time-averaged slope within $t=\pm 1$\,ps
shows the linear progression which corresponds to the quadratic
phase coefficient $b_2=\pm 25\,000$\,fs$^2$.  The major effect of
the quadratic phases on the photoassociation is a strong reduction
of the excited state population (and hence the molecular ion
signal). For both positive and negative chirps, the molecular ion
rate drops below 25\% of the zero-chirp level for
$b_2=\pm25\,000$\,fs$^2$. A strong reduction in ion rate supports
the argument that we are photoassociating to the 5s+5p asymptote
outside of the perturbative regime, since non-linear processes
require high peak intensities during the pulse maximum.

The spectral composition of the pulses, of course, is unaltered and
clearly the dynamics are similar to the unchirped case. The
modulations are observable even for large values of $b_2$,
decreasing only in amplitude whereas their period remains unchanged.
The main difference in the dynamics is a phase-shift between the
peak at $t=0$ and the oscillations, which is chirp-dependent.  This
is discussed in more detail in the accompanying theoretical paper
\cite{merli2009}.  The contrast between these results and the
proposed efficiency increase by chirped pulses, e.g. in
\cite{vala2001} arises from the fact that these proposals consider
resonant (within pulse spectrum) excitation to bound levels in the
free-bound transitions. As discussed above this is not the case in
our photoassociation experiments which mainly rely on non-linear,
off-resonant excitations.

\subsection{Fluorescence and trap loss}

\begin{figure}
\begin{center}
\includegraphics[width=\columnwidth]{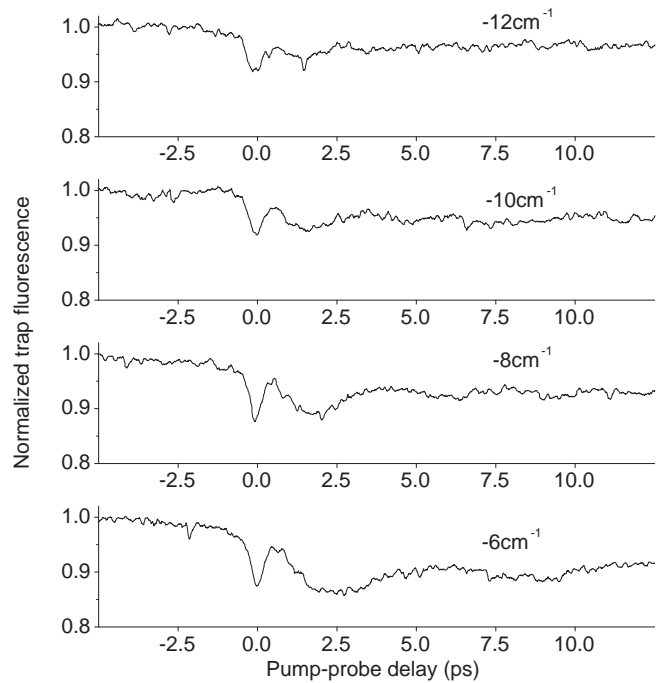}
\caption {Pump-Probe signals of trap fluorescence under variation of
  pump-probe delay using a D1 pulse.} \label{fig11_trap_fluo}
\end{center}
\end{figure}

\begin{figure}
\begin{center}
\includegraphics[width=\columnwidth]{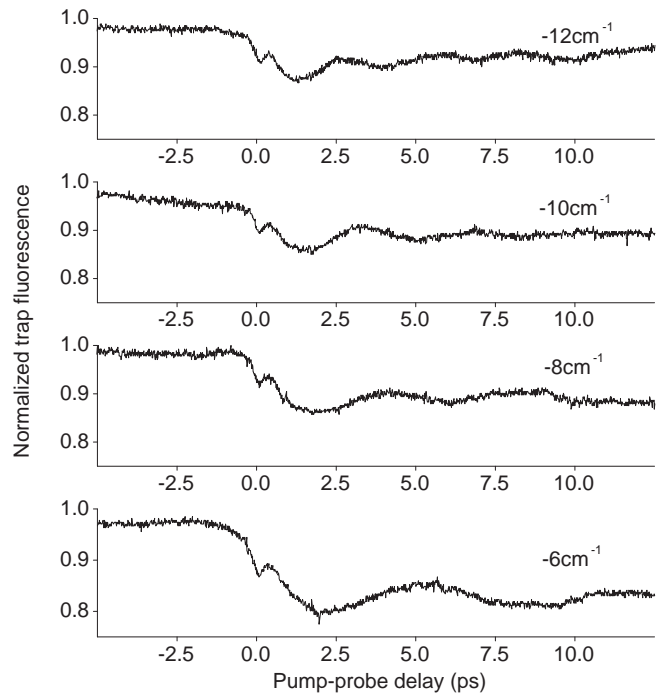}
\caption {Pump-Probe signals of trap fluorescence under variation of
  pump-probe delay using a D2 pulse.} \label{fig11_trap_fluo_D2}
\end{center}
\end{figure}

Similar dynamics are found for the atomic-loss in the MOT as for the
photoassociation of free-atom pairs. This is detected via the variation of the
trap fluorescence during the pump-probe scans, which indicates a
delay-dependent loss of atoms from the trap (excitation and loss processes
will be discussed shortly).  As shown in figures \ref{fig11_trap_fluo} and
\ref{fig11_trap_fluo_D2} the atomic losses increase for smaller cut-off
detunings and show the characteristic modulations. The results look closely
related to those found in \cite{monmayrant2006}.

In the present experiments, the trap-loss rate at t$>0$, due to the
femtosecond lasers, is estimated to be about $4\cdot10^6$ atoms/s, when using
a D1 pulse, which exceeds the femtosecond photoassociation signal by three
orders of magnitude.  Similarly to the collision pairs, single atoms are
off-resonantly excited to the 5p$_{1/2}$ or 5p$_{3/2}$ state by a
pump-pulse. Trap loss occurs by the subsequent probe-pulse excitation. The
probe pulse spectrum reaches from low lying Rydberg states at n=8 up to the
ionization threshold. Thus atoms may either be directly ionized or excited to
Rydberg states, the decay of which also results in trap loss.

The modulation on the trap fluorescence shows that large numbers of atoms
interact with the pump pulse field, in the same manner as discussed for the
photoassociated molecules. The large loss rate of atoms as compared to the
molecular ion count rate shows that atomic excitation is by far the dominant
process occurring in the trap during the pump-probe experiments.  With this in
mind, the femtosecond photoassociation discussed in the previous sections may
also be regarded as the excitation of single atoms plus the interaction with a
neighboring atom. Only a small fraction of atoms which have a close partner in
the interaction volume can be excited to the molecular ionic state by the pair
of pump-probe pulses. Due to the mass selection by the rf quadrupole, the ion
detection is restricted to molecular ions and thereby selects this small
fraction of close pairs in the interaction volume.

\section{Conclusion}

In summary, we have demonstrated the photoassociation of ultracold atoms using
shaped ultrashort laser pulses. The femtosecond laser mainly excites into
weakly bound excited molecular states outside the spectral profile of the
pulses. This is due to the non-perturbative coupling between the light field
and the ultracold atoms \cite{merli2009}. In order to excite into more deeply
bound states the density of the atomic sample, and thereby the Franck-Condon
overlap between the scattering wavefunction and the excited molecular state,
has to be significantly increased.

The use of a spectral cut to minimize spectral components near the
atomic resonance has aided us in the experiment in two ways:
Firstly, trap-loss due to ionization and radiation pressure was
strongly suppressed. Secondly, coherent dynamics of the newly
photoassociated molecules was revealed.  By applying a linear chirp
on the pump pulse we have been able to alter these coherent
dynamics. The periodic oscillations can be attributed to coherent
transient dynamics \cite{merli2009}. In addition, we see weak
evidence for molecules formed by the ultrashort pulse, which have
subsequently spontaneously decayed into the ground electronic state.

In order to observe vibrational wavepacket motion, ultrafast laser
pulse control in the picosecond regime is needed
\cite{monmayrant2004:rsi}, which fit to the time scale of
vibrational dynamics at large internuclear distance. Combined with
the high density this should allow to efficiently prepare deeply
bound molecules in a pump-dump scheme.

\section{Acknowledgments}
This work was supported by the DFG within the framework of the SFB 450 and the
SPP 1116. S. G. acknowledges support from the Studienstiftung des Deutsches
Volkes.

\end{document}